\documentstyle[sprocl,epsfig]{article}

\def\be{\begin{equation}}
\def\ee{\end{equation}}
\def\bea{\begin{eqnarray}}
\def\eea{\end{eqnarray}}

\newcommand{\case}[2]{\mbox{\footnotesize $\displaystyle \frac{#1}{#2}$}}
\newcommand{\lsim}{\mathrel{\rlap{\lower4pt\hbox{\hskip0pt$\sim$}}
\raise2pt\hbox{$<$}}}
\newcommand{\gsim}{\mathrel{\rlap{\lower4pt\hbox{\hskip0pt$\sim$}}
\raise2pt\hbox{$>$}}}

\begin{document}

\title{DSE HADRON PHENOMENOLOGY}

\author{M.\ B.\ HECHT, C.\ D.\ ROBERTS and S.\ M.\ SCHMIDT}

\address{Physics Division, Bldg. 203, Argonne National Laboratory\\
Argonne IL 60439-4843, USA}


\maketitle
\abstracts{A perspective on the contemporary use of Dyson-Schwinger
equations, focusing on some recent phenomenological applications: a
description and unification of light-meson observables using a one-parameter
model of the effective quark-quark interaction, and studies of leptonic and
nonleptonic nucleon form factors.\\[1ex]
%
}\vspace*{-2ex}

\hspace*{-\parindent}{\bf 1~~~Introduction.}~~The theory and phenomenological
application of Dyson-Schwinger equations (DSEs) have seen something of a
renaissance.  For example, they have been applied simultaneously to phenomena
as apparently unconnected as low-energy $\pi \pi$ scattering,\cite{pipi} $B
\to D^\ast$ decays$\,$\cite{mishaSVY} and the equation of state for a quark
gluon plasma;\cite{bastiscm} and there are renewed attempts$\,$\cite{gluonIR}
to understand the origin of the infrared enhancement
necessary$\,$\cite{fredIR} in the kernel of the quark DSE (QCD gap equation)
to generate dynamical chiral symmetry breaking (DCSB).  Also significant is
the appreciation$\,$\cite{bandocdrpionmrpion} that in this approach current
algebra's anomalies remain a feature of the global aspects of DCSB.
\medskip

\hspace*{-\parindent}{\bf 2~~~Meson Observables.}~~The renormalised gap
equation in Refs.~[\ref{mr97R}-\ref{mtkaonR}] is
\begin{equation}
S(p)^{-1} = Z_2 \,(i\gamma\cdot p + m_{\rm bare})
+ \int^\Lambda_q \, {\cal G}((p-q)^2)\, D_{\mu\nu}^{\rm free}(p-q)
\frac{\lambda^a}{2}\gamma_\mu S(q) \frac{\lambda^a}{2}\gamma_\nu \,,
\end{equation}
where $\int^\Lambda_q := \int^\Lambda d^4 q/(2\pi)^4$ represents mnemonically
a {\em translationally-invariant} regularisation of the integral, with
$\Lambda$ the regularisation mass-scale.  $Z_2$ is the quark wave function
renormalisation constant, which depends on the $\Lambda$ and the
renormalisation point, $\zeta$, and the renormalised current-quark mass is
\begin{equation}
\label{mzeta}
m(\zeta) := m_{\rm bare}(\Lambda)/Z_m(\zeta^2,\Lambda^2);\;\;
Z_m(\zeta^2,\Lambda^2) = Z_4(\zeta^2,\Lambda^2)/Z_2(\zeta^2,\Lambda^2)\,,
\end{equation}
where $Z_4$ is the renormalisation constant for the mass-term in the QCD
action.  

The model is specified by a choice for the effective interaction
\begin{equation}
\label{gk2VM}
\frac{{\cal G}(k^2)}{k^2} = \frac{4\pi^2}{\omega^6} D k^2 {\rm
e}^{-k^2/\omega^2} + 4\pi\,\frac{ \gamma_m \pi} {\case{1}{2} \ln\left[\tau +
\left(1 + k^2/\Lambda_{\rm QCD}^2\right)^2\right]} {\cal F}(k^2) \,,
\end{equation}
with ${\cal F}(k^2)= [1 - \exp(-k^2/[4 m_t^2])]/k^2$, $\gamma_m= 12/25$,
$\Lambda_{\rm QCD}=0.234\,$GeV, and fixed values of
$\omega=0.3\,$GeV$(=1/[.66\,{\rm fm}])$ and $m_t=0.5\,$GeV$(=1/[.39\,{\rm
fm}])$.  The sole parameter is the mass-scale: $D$.

The qualitative features of Eq.~(\ref{gk2VM}) are clear: the first term
provides for strength in the infrared that is known to be necessary to
support DCSB; the second term is proportional to the one-loop QCD
running-coupling at large spacelike-$k^2$, has no singularity on the
real-$k^2$ axis, and ensures that calculated quantities exhibit the one-loop
renormalisation group flow of QCD.  The latter characteristic follows
necessarily from the fact that a weak-coupling expansion of the DSEs yields
all the diagrams of perturbation theory.  This limits model-dependence to the
infrared.  Once a truncation of the quark DSE is specified, the form of the
meson Bethe-Salpeter equation (BSE) follows immediately using the systematic
Ward-Takahashi identity preserving procedure of Ref.~[\ref{axelR}].

The single model parameter, and the $u=d$ and $s$ current-quark masses were
varied$\,$\cite{mr97,pieterVM} so as to obtain a good description of
low-energy $\pi$- and $K$-meson properties, using a renormalisation point
$\zeta=19\,$GeV that is large enough to be in the perturbative domain.  The
fitting requires the repeated solving of the quark DSEs and meson BSEs.  It
could self-consistently yield $D\equiv 0$, which would indicate that
agreement with observable phenomena precludes an infrared enhancement in the
effective interaction.  However, that was not the case and a good fit
required
\begin{equation}
\label{paramsMT}
\begin{array}{ccc}
D= (1.12\,{\rm GeV})^2\; & \mbox{and}\;\;m_{u,d}^{1\,{\rm GeV}} =
5.5\,{\rm MeV}\,,\; & m_s^{1\,{\rm GeV}} = 124\,{\rm MeV}\,,
\end{array}
\end{equation}
and yields the results in Table~\ref{tableMesons}, which are characterised by
a root-mean-square error over predicted quantities of just $3.6\,$\%.  The
qualitative features of the dressed-quark propagator obtained in these
studies have recently been confirmed in lattice
simulations.\cite{tonylatticequark} 

\begin{table}[t]
\begin{center}
\parbox{33em}{\caption{Masses and decay constants [in GeV] of light vector
and flavour nonsinglet pseudoscalar mesons calculated using a
renormalisation-group-improved rainbow-ladder truncation.  The underlined
quantities were fitted.  The ``Obs.'' values of the masses are taken from
Ref.~[\protect\ref{pdg98R}], as are $f_\pi$, $f_K$.  The analogous vector
meson decay constants are discussed in
Refs.~[\protect\ref{mishaSVYR},\protect\ref{pieterVMR}] (Adapted from
Ref.~[\protect\ref{pieterVMR}].)\hspace*{\fill}
\label{tableMesons}}}
\begin{small}
\[
\begin{array}{l|ccccccccccc}
        & -(\langle \bar q q \rangle^0_{1\,{\rm GeV}})^{1/3} &
        m_\pi & m_K & m_\rho & m_{K^\ast} & m_\phi & f_\pi & f_K & f_\rho &
        f_{K^\ast} & f_\phi \\\hline
\rule{0mm}{1.2em}{\rm Obs.} & 0.236\,\mbox{\cite{derek}}
         & 0.139 & 0.496 & 0.770 & 0.892 & 1.020 &
        0.130 & 0.160 & 0.216 & 0.225 & 0.238\\
\rule{0mm}{1.2em}{\rm Calc.} & 0.242 & \underline{0.139} & \underline{0.496}
        & 0.747 & 0.956 & 1.088 & \underline{0.130} & 0.154 & 0.197 & 0.246 &
        0.255 \\\hline
\end{array}
\]
\end{small}
\vspace*{-1.5em}

\end{center}
\end{table}

\begin{figure}[t]
\centering{\
\epsfig{figure=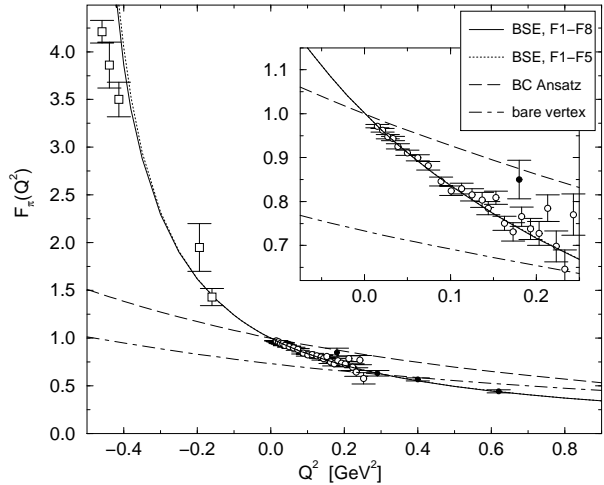,height=6.5cm}}\vspace*{-2ex}

\parbox{33em}{\caption{Pion form factor calculated directly from the
interaction in Eq.~(\protect\ref{gk2VM}): solid line.  The data are from
Refs.~[\protect\ref{piondatR}].  No parameters were varied to obtain the
result.  The other curves depict results from simplified calculations; in
particular, the dash-dot line uses $\Gamma_{\mu}^\gamma(k;P) = \gamma_\mu$,
which violates the Ward-Takahashi identity.  (Adapted from
Ref.~[\protect\ref{mtpionR}].)\hspace*{\fill}\label{mtpionff}}}\vspace*{-3ex}
\end{figure}

The same model can be employed without modification in calculating the
impulse approximation to the light pseudoscalar meson form
factors.\cite{mtpion,mtkaon} In this application one needs additionally to
solve the inhomogeneous vector vertex equation, which describes the
dressed-quark--photon coupling {\it and} exhibits a pole at $P^2+m_V^2=0$;
i.e., at the particular flavour channel's vector meson mass.  It is only by
solving the inhomogeneous vertex equation that one can unambiguously evolve
the form factor from the spacelike into the timelike region and vice versa.

The result for the pion form factor is illustrated in Fig.~\ref{mtpionff},
with $r_\pi^2= (0.67\,{\rm fm})^2$ cf.\ expt.\cite{piondat} $(0.663\pm
0.007\,{\rm fm})^2$.  Good results are also obtained for the charged and
neutral kaon form factors, with $r_{K^+}^2= (0.62\,{\rm fm})^2$ and
$r_{K^0}^2= -(0.29\,{\rm fm})^2$ cf.\ expt.\cite{piondat,kaondat} $r_{K^+}^2=
(0.583 \pm 0.043\,{\rm fm})^2$ and $r_{K^0}^2= -(0.23\pm 0.06\,{\rm fm})^2$.
These results are {\it too} good because the calculations neglect
meson-rescattering contributions that, in all channels, are additive in
magnitude, with corrections of up to 15\%.\cite{pionloops} Nevertheless they
are a significant step, providing a manifestly Poincar\'e invariant
calculation and unification of light-meson observables, with the particular
feature that all calculated quantities are independent of the definition of
the relative momentum, which is arbitrary in a covariant formulation.  A
three-dimensional reduction of the Bethe-Salpeter equation is unnecessary and
is not employed.  Furthermore, current conservation is automatic and the
neutral mesons are {\it neutral} without fine-tuning.  The same framework
predicts:\cite{mrpion} $Q^2 F_{\rm meson}(Q^2) =\,$constant, as
$Q^2\to\infty$, in accordance with the pQCD expectation, which is only
possible because the pseudoscalar meson Bethe-Salpeter amplitude necessarily
has pseudovector components.\cite{mrt98}
\medskip

\hspace*{-\parindent}{\bf 3~~~Nucleon
Observables.}~~Reference~[\ref{reinhardR}] is an extensive study of the octet
and decuplet baryon spectrum based on a quark-diquark Fadde$^\prime$ev
equation.  It represents the nucleon as a composite of a quark and pointlike
diquark, which are bound together by a repeated exchange of roles between the
dormant and diquark-participant quarks, and demonstrates conclusively that an
accurate description of the spectrum is possible using these degrees of
freedom.  This motivates and supports the product {\it Ansatz} for the
nucleon's amplitude used in Refs.~[\ref{jcrb1R},\ref{jcrb2R}] to calculate a
wide range of leptonic and nonleptonic nucleon form factors:
\begin{equation}
\label{Psi}
\Psi(p_i;\alpha_i,\tau_i;\alpha,\tau)= 
\delta^{\tau \tau_3}\,
%
\psi_{\alpha\alpha_3}(p_1+p_2,p_3)
\Delta(p_1+p_2)\, \Gamma_{\alpha_1 \alpha_2}^{\tau_1
\tau_2}(p_1,p_2) \,,
\end{equation}
with the nucleon's momentum: $P=p_1+p_2+p_3$, $P^2=-M^2$.  In
Eq.~(\ref{Psi}), $(\alpha_i,\tau_i)$ are quark spinor and isospin labels, and
$(\alpha,\tau)$ are those of the nucleon, and
\begin{equation}
\psi_{\alpha\alpha_3}(\ell_1,\ell_2)
= \delta_{\alpha \alpha_3} \psi_1(\ell^2)
- \frac{1}{M}\left( i\gamma\cdot \ell - \ell\cdot \hat P \delta_{\alpha
        \alpha_3}\right) \psi_2(\ell^2)\,,
\end{equation}
with $\ell = (1/3)\ell_1 - (2/3)\ell_2$ and $\hat P^2 = -1$, is a
Bethe-Salpeter-like amplitude characterising the correlation between the
quark and the diquark.  In this form, $\psi_1$ describes the upper-component
of the positive-energy nucleon spinor and $\psi_2$ the
lower-component.\cite{piller} In addition, $\Delta(K)$ describes the
pseudo-particle propagation characteristics of the diquark, and
\begin{eqnarray}
\label{gdq}
\Gamma_{\alpha_1 \alpha_2}^{\tau_1 \tau_2}(p_1,p_2) & = &
(C i\gamma_5)_{\alpha_1 \alpha_2}\, (i\tau^2)^{\tau_1\tau_2}\,
\Gamma (\ell^2)\,,\;\ell = \case{1}{2}q_1 - \case{1}{2} q_2\,,
\end{eqnarray}
represents the momentum-dependence, and spin and isospin character of the
diquark correlation; i.e., it corresponds to a Bethe-Salpeter-like amplitude
for what here is a nonpointlike diquark.  

Equations~(\ref{Psi})--(\ref{gdq}) describe the nucleon as a composite of a
quark and a scalar-diquark correlation, and in
Refs.~[\ref{jcrb1R},\ref{jcrb2R}] the scalar functions were parametrised:
\begin{equation}
\label{littlepsi}
\begin{array}{ccc}
\psi_1(\ell) = \frac{1}{{\cal N}_\Psi}\,{\cal F}(\ell^2/\omega_\psi^2)\,,\;&
\Gamma(\ell) = \frac{1}{{\cal N}_\Gamma}\, {\cal F}(q^2/\omega_\Gamma^2)\,,\;&
\Delta(K)   =  \frac{1}{m_\Delta^2}\,{\cal F}(K^2/\omega_\Gamma^2)\,,
\end{array}
\end{equation}
${\cal F}(y) = (1- {\rm e}^{-y})/y$, $\psi_2=0$, with ${\cal N}_\Psi$ and
${\cal N}_\Gamma$ being the calculated nucleon and $(ud)$-diquark
normalisation constants, which ensure composite electric charges of $1$ for
the proton and $1/3$ for the diquark.  The parameters $\omega_\psi=0.2\,$GeV,
$\omega_\Gamma=1.4\,$GeV and $m_\Delta=0.63\,$GeV were fixed$\,$\cite{jcrb2}
in a least-squares fit to the proton's charge form factor on
$Q^2\in[0,3]\,$GeV$^2$.  A good description is obtained and the parameter
values demonstrate the internal consistency of the model.  $d_\Gamma:=
1/\omega_\Gamma$ is a measure of the mean separation between the quarks
constituting the scalar diquark and $d_\psi:= 1/\omega_\psi$ is the analogue
for the quark-diquark separation.  $d_\Gamma<d_\psi$ is necessary if the
quark-quark clustering interpretation is to be valid.  $\ell_{(ud)_{0^+}}=
1/m_\Delta$ is a measure of the range over which the diquark may propagate
and that must be significantly less than the nucleon's diameter.

One particular highlight of the calculations is the result that the
nonpointlike nature of the diquark allows a better description of the
nucleons' magnetic form factors than is possible in pointlike diquark models
and, importantly, $|\mu_n/\mu_p|=0.55$ cf.\ expt.\ 0.68, whereas pointlike
scalar-diquark models always yield $|\mu_n/\mu_p|<0.5$.  Another is a
prediction for the ratio $\mu_p G_E^p(q^2)/G_M^p(q^2)$ that is in
semi-quantitative agreement with recent results from TJNAF.\cite{HallA} A
defect is that $|r_n^2|$ is 60\% too large.

\begin{figure}[t]
\centering{\
\epsfig{figure=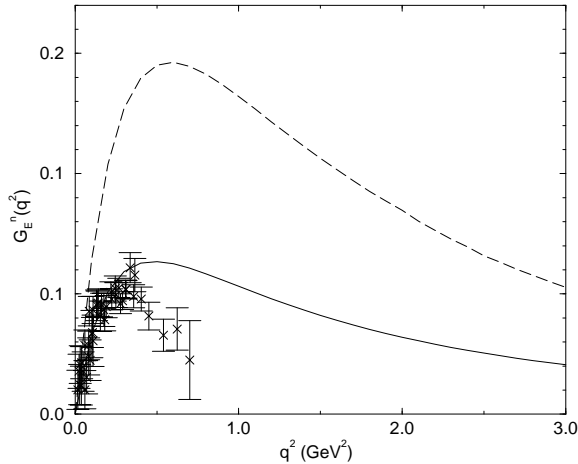,height=6.2cm}}
\parbox{33em}{\caption{\label{nnff} Solid line: neutron electric form factor
calculated with $\psi_2=\psi_1$; i.e., including the lower component of the
neutron spinor.  Dashed line: the result with $\psi_2=0$.\protect\cite{jcrb1}
Data from Ref.~[\protect\ref{saclayR}], extracted using the Argonne V18
potential.\protect\cite{bobpot} Improving the product {\it Ansatz},
Eq.~(\protect\ref{Psi}), by including $\psi_2$ significantly improves the
estimate of $G_E^n$.  However, this preliminary result is too good as we
still ignore the axial vector diquark correlation.\hspace*{\fill}}}\vspace*{-2ex}
\end{figure}

There are two obvious improvements -- include: the lower component of the
nucleon spinor, $\psi_2\neq 0$; and the axial-vector diquark.  The first is
underway and we report preliminary results for $\psi_2=\psi_1$.  A fit to
$G_E^p(p^2)$ now requires
\begin{equation}
\label{paramsNFF}
\begin{array}{l|ccc}
               & \;\omega_\psi \;& \;\omega_\Gamma \;&\; m_\Delta \\\hline
\mbox{in GeV}  &  \;0.19\;  &\; 0.68 \;  &\; 0.64
\end{array}\,,\;\;\;\;
\begin{array}{l|ccc}
               & \;1/\omega_\psi\; &\; 1/\omega_\Gamma\; &\; 1/m_\Delta \\\hline
\mbox{in fm}  &  1.03  \; &\; 0.29\; &\; 0.31
\end{array}\,;
\end{equation}
but this represents only a small change in the fitting parameters that
preserves the model's internal consistency [cf.\ after Eq.~(\ref{littlepsi})].
These values yield
\begin{equation}
\label{results}
\begin{array}{c|ccccc}
          & \;r_p^2\,({\rm fm}^2)\; & \;r_n^2\,({\rm fm}^2)\;&
          \;\mu_p (\mu_N)\; &\; \mu_n (\mu_N)\; &\;\mu_n/\mu_p \\\hline
{\rm Emp.} & (0.87)^2 & -(0.34)^2 & 2.79 & -1.91 & -0.68 \\
{\rm Calc.}& (0.78)^2 & -(0.33)^2 & 2.81 & -1.61 & -0.57 \\ \hline
\mbox{{\rm Old~Calc.\cite{jcrb1}}}
           & (0.79)^2 & -(0.43)^2 & 2.88 & -1.58 & -0.55\\
\end{array}
\end{equation}
and the recalculated neutron electric form factor in Fig.~\ref{nnff}.
As elucidated in Refs.~[\ref{jcrb1R}], of these calculated quantities only
$G_E^n$ involves a cancellation between contributions from the five diagrams
that constitute the impulse approximation to the nucleon's electromagnetic
current in this model, and only observables tied to this form factor are
dramatically affected by the improvement of the product {\it Ansatz},
Eq.~(\ref{Psi}).  We anticipate that this is a general feature; i.e., only
observables receiving interfering contributions; e.g., $g_A$ and the
isoscalar tensor coupling $f_{\omega NN}$, are significantly modified by
improvements.  If this were not the case, simple models could not be
generally efficacious.
\medskip

\hspace*{-\parindent}{\bf Epilogue.}~~This overview is necessarily brief.  It
does no more than point to recent successes and says nothing of contemporary
challenges.  One such is to comprehend the origin of the infrared enhancement
in the kernel of the QCD gap equation that is necessary to ensure DCSB.
Another is to develop a BSE based understanding of the light scalar mesons
and the $\eta$-$\eta^\prime$ mass-splitting.  These aspects and more are
canvassed in Ref.~[\ref{bastirevR}].
\medskip

\hspace*{-\parindent}{\bf Acknowledgments.}~~CDR is grateful to the staff of
the Special Research Centre for the Subatomic Structure of Matter at the
University of Adelaide for their hospitality and support during this
workshop, and in the two preceding weeks; and we acknowledge helpful
communications with J.C.R.~Bloch.  This work was supported by the US
Department of Energy, Nuclear Physics Division, under contract
no.\ W-31-109-ENG-38, and benefited from the resources of the National Energy
Research Scientific Computing Center. SMS is grateful for financial support
from the A.\ v.\ Humboldt foundation.

\begin{flushleft}

\end{flushleft} 
\end{document}